\documentclass[showpacs,preprintnumbers,superscriptaddress]{revtex4}
\usepackage{CJK}
\usepackage{amsmath,amssymb,graphicx,bm}
\begin{document}

\title{Horizon thermodynamics in $f(R)$ theory}

\author{Yaoguang Zheng}
%\email{hesoyam12456@163.com}
\affiliation{College of Physical Science and Technology, Hebei University, Baoding 071002, China}
\author{Rongjia Yang \footnote{Corresponding author}}
\email{yangrongjia@tsinghua.org.cn}
\affiliation{College of Physical Science and Technology, Hebei University, Baoding 071002, China}
\affiliation{Hebei Key Lab of Optic-Electronic Information and Materials, Hebei University, Baoding 071002, China}

\begin{abstract}
We investigate whether the new horizon first law proposed recently still work in $f(R)$ theory. We identify the entropy and the energy of black hole as quantities proportional to the corresponding value of integration, supported by the fact that the new horizon first law holds true as a consequence of equations of motion in $f(R)$ theories. The formulas for the entropy and energy of black hole found here are in agreement with the results obtained in literatures. For applications, some nontrivial black hole solutions in $f(R)$ theories have been considered, the entropies and the energies of black holes in these models are firstly computed, which may be useful for future researches.
\end{abstract}


\maketitle

\section{Introduction}
It is well-known that there is a profound connection between gravity and thermodynamics: the spacetime with horizons can be described by thermodynamic laws. Bekenstein found that the area of a black hole can be seen as its entropy \cite{Bekenstein:1973ur}. Four laws of black hole mechanics were proposed in \cite{Bardeen:1973gs}. It has been shown that the entropy of a black hole can be taken as the Noether charge associated with the diffeomorphism invariance of the theory of gravity \cite{Wald:1993nt, Iyer:1994ys}. From the first law of thermodynamics Einstein equation has been derived in \cite{Jacobson:1995ab}. Non-equilibrium thermodynamics of spacetime has been investigated in \cite{Eling:2006aw}. By using a more general definition of the Noether charge entropy, the equations of motion of generalized theories of gravity are equivalent to the thermodynamic relation $\delta Q=T\delta S$ \cite{Brustein:2009hy}. This attempt also has been considered in modified gravity theories: such as $f(R)$ theory \cite{Elizalde:2008pv}, Lancos-Lovelock gravity \cite{Paranjape:2006ca}, and the scalar-Gauss-Bonnet gravity \cite{Bamba:2009gq}. In \cite{Padmanabhan:2002sha}, a general formalism for understanding the thermodynamics of horizons in spherically symmetric spacetimes was developed. For stationary axis-symmetric horizons and time dependent evolving horizons, it has been shown that the near horizon structure of Einstein equations can be expressed as a thermodynamic identity under the virtual displacement of the horizon \cite{Kothawala:2007em}. It also has been shown that the gravitational field equations of $n+1$ -dimensional topological black holes with constant horizon curvature, in cubic and quartic quasi-topological gravity, can be recast in the form of the first law of thermodynamics \cite{Sheykhi:2014rka}. All these studies were based on some assumptions, such as horizon, null surfaces, Unruh temperature, and so on. In \cite{Yang:2011sx}, without assuming a temperature or a horizon the thermal entropy density has been obtained for any arbitrary spacetime, implying that gravity possesses thermal effects, or, thermal entropy density possesses effects of gravity. These results has been generalized to the case of nonzero chemical potential \cite{Yang:2014kna}.

When assuming a horizon equation of state, one can get a horizon first law by considering a virtual displacement, from which the entropy can be obtained \cite{Padmanabhan:2002sha}. Recently a new horizon first law was suggested in \cite{Hansen:2016gud}, in this approach both the entropy and the free energy are derived concepts, and from which the standard horizon first law is recovered by a Legendre projection. For Einstein gravity and Lovelock gravity which only give rise to second-order field equation for all metric components, their results have establish a way of how to formulate consistent black hole thermodynamic without conserved charges. Here we will investigate whether the new horizon first law still work in $f(R)$ gravity which has fourth-order field equation. In these higher order gravitational theories, the issue associated with the energy of black hole is problematic. Several attempts to find a satisfactory answer to this problem have been carried out (see for example \cite{Cognola:2011nj, Deser:2002jk, Deser:2007vs, Abreu:2010sc, Cai:2009qf} and references therein). We find that the new horizon first law can give not only the entropy but also the energy of black holes in $f(R)$ theories, which are consistent with the results obtained in literatures.

The rest of the paper is organized as follows. In Sec. II, we sketch the suggestion of the new horizon first law. In Sec. III, we consider whether the new horizon first law holds in $f(R)$ theory. In Sec. IV, we discuss applications for some $f(R)$ theories. Conclusions and discussions are presented in Sec. V.

\section{The new horizon first law}

When discussing the thermodynamics of gravitation, it is usual to assume that the sources of thermodynamic system are also the sources of gravity, which as a principle was firstly proposed in \cite{Yang:2014kna}, such as: $\rho, P|_{\rm{gravitational~ source}}=\rho, P|_{\rm{thermal~ source}}$, with $\rho$ the energy density and $P$ the pressure. For horizon thermodynamics, this assumption implies that the radial component of the stress-energy tensor serves as a thermodynamic pressure: $P=T^{r}_{~r}|_{r_{+}}$. Further one can assume an horizon equation of state, $P=P(V, T)$, where the temperature is identified with the Hawking temperature and the horizon is assigned a geometric volume $V=V(r_{+})$ \cite{Parikh:2005qs}. Then one can reasonably realize the radial Einstein equation. Considering a virtual displacement of the horizon \cite{Padmanabhan:2002sha}, the horizon equation of state can be reformulate as a horizon first law
\begin{eqnarray}
\label{HFL}
\delta E=T\delta S-P\delta V,
\end{eqnarray}
where $S$ is the horizon entropy and $E$ is quasilocal energy of the black hole. In theory of Einstein gravity, $E$ turns out to be the Misner-Sharp energy \cite{Misner:1964je}. It was shown that the horizon first law (\ref{HFL}) is a special case of the `unified first law' \cite{Hayward:1997jp}.

These results are quite inspiring, but there are some problems in this procedure that are needed to be further checked up. Firstly, the thermodynamic variables were derived that was vague in the original derivation, and these variables require further determination. The second problem is the restriction of the virtual displacements $\delta r_+$ of the horizon radius. This makes the horizon first law (\ref{HFL}) to be of `\emph{cohomogeneity-one}': as a function only depending on $r_+$, because both $S$ and $V$ arefunctions of only $r_+$. In fact, equation (\ref{HFL}) can be written in $\delta E=\left(TS^\prime+PV^\prime\right)\delta r_+$, where the primes represents the derivative with respected to $r_+$. This makes the terms `work' and `heat' unclear and results to a `vacuum interpretation' of the first law (\ref{HFL}) \cite{Hansen:2016gud}. To avoid these two problems, the key point is to vary the horizon equation of state with the temperature $T$ and the pressure $P$ as independent thermodynamic quantities. This procedure leads to a new horizon first law \cite{Hansen:2016gud}.
\begin{eqnarray}
\label{nhel}
\delta G=-S\delta T+V\delta P,
\end{eqnarray}
which is obviously of cohomogeneity-two (depending on both $T$ and $P$) and non-degenerate ($T$ and $P$ can vary independently). Furthermore, for specified volume, pressure, and temperature, the horizon entropy $S$ and the Gibbs free energy $G$ are derived concepts. The standard horizon first law (\ref{HFL}) can be derived though a degenerate Legendre transformation $E=G+TS-PV$. This new derivation indicates that horizon thermodynamics has practical utility and provides further evidence that gravitational field equations can indeed be interpreted as an equation of state.

We first briefly review this method to horizon thermodynamics in 4-dimensional Einstein gravity to explain how it works \cite{Hansen:2016gud}. Considering the spacetime of a static spherically symmetric black hole whose geometry is given by
\begin{eqnarray}
\label{metric}
ds^2=-B(r)dt^2+\frac{dr^2}{B(r)}+r^2d\Omega^2,
\end{eqnarray}
where the event horizon is located at $r=r_+$ which is the largest positive root of $B(r_+)=0$ which fulfils $B'(r_+)\neq 0$ . Supposing minimal coupling to the matter, with the stress-energy tensor $T_{\mu\nu}$, the radial Einstein equation yields
\begin{eqnarray}
\label{ER}
8\pi T^r_{\ r}|_{r_+}=G^r_{\ r}|_{r_+}=\frac{B'(r_+)}{r_+}-\frac{1-B(r_+)}{r^2_+}.
\end{eqnarray}
We take the units in which $G=c=\hbar=1$ throughout this paper. Assuming that the thermal sources are also the gravitational sources \cite{Yang:2014kna}, we have $P=T^r_{~r}$ and identify \cite{Padmanabhan:2002sha}
\begin{eqnarray}
\label{T}
T=\frac{B'(r_+)}{4\pi},
\end{eqnarray}
as the the temperature $T$. At the horizon, the radial Einstein equation (\ref{ER}) can be rewritten as
\begin{eqnarray}
\label{P}
P=\frac{T}{2r_{+}}-\frac{1}{8\pi r^2_{+}},
\end{eqnarray}
which is the horizon equation of state. The identification of the temperature $T$ in (\ref{T}) is via standard arguments in thermal quantum field theory; it does not fall back on any gravitational field equations \cite{Hansen:2016gud}. The definition the pressure in (\ref{P}), according to the conjecture proposed in \cite{Yang:2014kna}, is identified with the $(^r_{r})$ component of the matter stress-energy, also independent of any gravitational field equations. With this information, it is reasonable to suggest that the radial field equation for a gravitational theory under consideration takes the following form \cite{Hansen:2016gud}
\begin{eqnarray}
\label{7}
P=D(r_+)+C(r_+)T,
\end{eqnarray}
where $D$ and $C$ are some functions of $r_+$ that in general depend on the gravitational theory under consideration, like the linear equation of state in the temperature $T$. To obtain the new horizon first law from the generalized equation of state (\ref{7}), assuming a virtual displacements $\delta r_+$ of the horizon radius and varying the equation (\ref{7}), then multiplying the geometric volume $V(r_+)$, it is straightforward to get
\begin{eqnarray}
\label{8}
V\delta P=V\left(D'+C' T\right)\delta r_{+}+VC\delta T,
\end{eqnarray}
in which $VC$ is just the entropy and $V\left(D'+C' T\right)\delta r_{+}$ is the variation of the Gibbs free energy $G$, comparing with the equation (\ref{nhel}). It is now easy to rewrite this equation as \cite{Hansen:2016gud}
\begin{eqnarray}
\label{9}
V\delta P=S\delta T+\delta G,
\end{eqnarray}
where
\begin{eqnarray}
\label{10}
G&=&\int V(r_+)D'(r_+)\,dr_{+}+T\int V(r_+)C'(r_+)dr_+ \nonumber\\
&=&PV-ST-\int V'(r_+)D(r_+)dr_+,
\end{eqnarray}
%%%%%%%%%%%
and
\begin{eqnarray}
\label{11}
S=\int V'(r_+)C(r_+)dr_+,
\end{eqnarray}
by making use of the integration by parts. The derivation of the equation (\ref{9}) depends only on the generalized equation of state with the form (\ref{7}), having not used the specific form of the volume. Using the degenerate Legendre transformation $E=G+TS-PV$, we obtain the energy as \cite{Hansen:2016gud}
\begin{eqnarray}
\label{energy}
E=-\int V'(r_+)D(r_+)dr_+.
\end{eqnarray}
Taking variation of the equation (\ref{energy}) and combining the equation (\ref{9}), one can obtain the horizon first law (\ref{HFL}) which is a special case of the `unified first law' \cite{Hayward:1997jp}.

Supposing that $P$, $T$, and $V$ can be identified as the pressure, the temperature, and the volume, then one can come to the conclusion that $S$, $G$, and $E$ are the entropy, the Gibbs free energy, and the energy of the black hole, respectively. For Einstein gravity in four dimensions, it is straightforward to have $D(r_+)=-(8\pi r^2_+)^{-1}$ and $C(r_+)=1/(2r_+)$ from (\ref{P}), yielding $S=\pi r^2_+$ from (\ref{11}) and $E=r_+/2$ from (\ref{energy}).
\section{The entropy and energy of black holes in $f(R)$ Theory}
The new horizon first law works well in Einstein's theory, as shown in the previous section. A question naturally rises whether it still work in other modified gravity theories? Here we consider this problem in $f(R)$ theories. The general action of $f(R)$ gravity theories in four-dimensional with source is
\begin{eqnarray}
\label{12}
I=\int \text{d}^{4}x\sqrt{-g}\left[\frac{f(R)}{2k^2}+L_{\rm m}\right],
\end{eqnarray}
where $k^2=8\pi$, $f(R)$ is a general function of the Ricci scalar $R$, and $L_{\rm m}$ is the matter Lagrangian. For physical meaning, function $f(R)$ must satisfy the stability conditions \cite{Pogosian:2007sw}: (a) no ghosts, $df/dR>0$; (b) no tachyons, $d^2f/dR^2>0$ \cite{Dolgov:2003px}; (c) $\lim_{R\rightarrow \infty} (f(R)-R)/R=0$ for the existence of an effective cosmological constant at high curvature; (d) $\lim_{R\rightarrow 0} d(f(R)-R)/dR=0$ for recovering general relativity at early time (allowing for vaccum solutions). Moreover, in order to give rise to stable solutions, $f(R)$ also must fulfill additional condition: $df/dR/d^2f/dR^2>R$ \cite{Sawicki:2007tf}. From the variation of the action (\ref{12}) with respect to the metric, the gravitational field equations are obtained as follows
\begin{eqnarray}
\label{13}
G_{\mu\nu}\equiv R_{\mu\nu}-\frac{1}{2}g_{\mu\nu}R=k^2\left(\frac{1}{F}T_{\mu\nu}+\frac{1}{k^2}\mathcal{T}_{\mu\nu} \right),
\end{eqnarray}
where $T_{\mu\nu}=\frac{-2}{\sqrt{-g}}\frac{\delta L_{\rm m}}{\delta g^{\mu\nu}}$ is the energy-momentum tensor of the matter,  $F=\frac{df}{dR}$, and $\mathcal{T}_{\mu\nu}$ is the tress-energy tensor of the effective curvature fluid which is given by
\begin{eqnarray}
\label{rtensor}
\mathcal{T}_{\mu\nu}=\frac{1}{F(R)}\left[\frac{1}{2}g_{\mu\nu}(f-RF)+\nabla_{\mu}\nabla_{\nu}F-g_{\mu\nu}\Box F \right],
\end{eqnarray}
where $\Box =\nabla ^{\gamma }\nabla _{\gamma}$. Using the relations $\Box F=\frac{1}{\sqrt{-g}}\partial_{\mu}[\sqrt{-g}g^{\mu\nu}\partial_{\nu}F]$ and assuming the geometry of a static spherically symmetric black hole takes the form of the equation (\ref{metric}), we derive after some calculations the $(^1_1)$ components of the Einstein tensor and the stress-energy tensor of the effective curvature fluid, respectively, as
\begin{eqnarray}
\label{gtensor11}
G^{1}_{1}=\frac{1}{r^2}(-1+rB'+B),
\end{eqnarray}
and
\begin{eqnarray}
\label{rtensor11}
\mathcal{T}^{1}_{1}=\frac{1}{F}\left[\frac{1}{2}(f-RF)-\frac{1}{2}B'F'-\frac{2}{r}BF'\right].
\end{eqnarray}
where the primes represents the derivative with respected to $r$. Taking the trace of the equation (13), we derive the relation
\begin{eqnarray}
\label{trace}
RF(R)-2f(R)+3\Box F(R)=k^2T.
\end{eqnarray}
Inserting equations (\ref{gtensor11}) and (\ref{rtensor11}) and $T^1_1=P$ into the equation (\ref{13}), we obtain after some appropriate arrangements
\begin{eqnarray}
\label{26}
k^2P=-\left[\frac{F}{r^2}+\frac{1}{2}(f-RF)\right]+\left(\frac{F}{r}+\frac{1}{2}F'\right)B'+\frac{FB}{r^2}+\frac{2BF'}{r}.
\end{eqnarray}
At the horizon, $r=r_+$, we have $B(r_+)=0$ and Hawking temperature $T=B'(r_+)/4\pi$, the equation (\ref{26}) reduces to
\begin{eqnarray}
\label{27}
P=-\frac{1}{8\pi}\left[\frac{F}{r^2_+}+\frac{1}{2}(f-RF)\right]+\frac{1}{4}\left(\frac{2F}{r_+}+F'\right)T.
\end{eqnarray}
Comparing Eq. (\ref{7}) with Eq. (\ref{27}), we then have
\begin{eqnarray}
\label{28}
D(r_+)=-\frac{1}{8\pi}\left[\frac{F}{r^2_+}+\frac{1}{2}(f-RF)\right],
\end{eqnarray}
and
\begin{eqnarray}
\label{29}
C(r_+)=\frac{1}{4}\left(\frac{2F}{r_+}+F'\right).
\end{eqnarray}
The geometric volume $V$ of the black hole in four-dimensional spacetime is $V(r_+)=4\pi r_+^3/3$. Inserting this equation and (\ref{29}) into Eq. (\ref{11}), the entropy is given by
\begin{eqnarray}
\label{31}
S&=&\int V'(r_+)C(r_+)dr_+=\int (2\pi r_{+}F+\pi r^2_{+}F')dr_+\\\nonumber
&=&F(r_+)\frac{A}{4},
\end{eqnarray}
which is $F(r_+)$ times the Bekenstein-Hawking Entropy, it is in agreement with results obtained by using the Euclidean semiclassical approach or the Wald entropy formula \cite{Dyer:2008hb, Vollick:2007fh, Iyer:1995kg}. By implementing the degenerate Legendre transformation $E=G+TS-PV$ and substituting the geometric volume $V$ and Eq. (\ref{28}) into Eq. (\ref{energy}), then we get the energy as
\begin{eqnarray}
\label{energy1}
E&=&-\int V'(r_+)D(r_+)dr_+\nonumber \\
&=&\frac{1}{2}\int \left[\frac{F}{r^2_+}+\frac{1}{2}(f-RF)\right]r^2_{+}dr_+,
\end{eqnarray}
which is also consistent with the expression obtained in \cite{Cognola:2011nj} where the entropy was pre-given as that obtained by using the Wald method. The results obtained here show that the new horizon first law is still respected by $f(R)$ theories.
\section{Applications}
In this section, we will illustrate the method with explicit examples of entropy and energy calculation by using the equations (\ref{31}) and (\ref{energy1}). We will analyze some $f(R)$ theories which have received a lot of attentions. These models admit solutions with constant Ricci curvature (such as Schwarzschild or the Schwarzschild-de Sitter solutions) or solutions with non-constant Ricci curvature.
\subsection{The constant Ricci curvature case}
As a simple but important example, we firstly consider a spherically symmetric solution of (\ref{13}) with constant curvature $R_0$, such as the Schwarzschild ($R_0=0$) or the Schwarzschild-de Sitter solution, in which $B(r)$ takes the form $B(r)=1-2M/r-R_{0}r^2/12$ with $M$ the mass of the black hole \cite{Multamaki:2006zb}. In this case, the trace equation (\ref{trace}) simply reduces to
\begin{eqnarray}
\label{tcond}
R_{0}F(R_0)=2f(R_0),
\end{eqnarray}
which must be fulfilled by any $f(R)$ theories allowing a Schwarzschild-de Sitter solution. At the horizon, one has $B(r_+)=0$ which gives
\begin{eqnarray}
\label{hc}
2M=r_{+}-\frac{1}{12}R_{0}r^3_{+}.
\end{eqnarray}
which relates the mass $M$, the radius $r_+$, and the Ricci curvature $R_0$ together.

The first one we focus on, introduced in \cite{Carroll:2003wy, Capozziello:2003gx} to give rise to
a late-time acceleration and reconsidered in \cite{Li:2007xn, Amendola:2006we}, is given by
\begin{eqnarray}
f(R) = R-\alpha R^n.
\end{eqnarray}
where $\alpha$ and $n$ are constant with $\alpha>0$ and $0<n<1$. Since $f(0)$=0, this model allows for a Schwarzschild solution. The condition to admit a Schwarzschild-de Sitter black hole, (\ref{tcond}), leads to $R_0=\{1/[(2-n)\alpha]\}^{1/(n-1)}$. From the equation (\ref{31}), we obtain the entropy of Schwarzschild-de Sitter black hole as
\begin{eqnarray}
\label{fs1}
S=\left(1-\alpha n R_0^{n-1}\right)\pi r^2_+=\frac{2-2n}{2-n}\frac{A}{4},
\end{eqnarray}
which is $\frac{2-2n}{2-n}$ times the Bekenstein-Hawking Entropy. Since $\alpha> 0$, we have $S\rightarrow A/4$ for $n\rightarrow 0$ and $S\rightarrow 0$ for $n\rightarrow 1$, meaning the entropy of Schwarzschild-de Sitter black hole in this $f(R)$ model is less than the Bekenstein-Hawking Entropy. While for Schwarzschild black hole, the entropy in this $f(R)$ theory is equivalent to Bekenstein-Hawking Entropy. The energy of the black hole is obtained form (\ref{energy1}) as
\begin{eqnarray}
E=\frac{1-n}{2-n} r_{+}-\frac{1-n}{12(2-n)} R_{0}r^3_{+}=\frac{2-2n}{2-n}M,
\end{eqnarray}
having the same limit properties as that in the entropy (\ref{fs1}), comparing with the result in Einstein's gravity. Since $0<n<1$, both the entropy and the energy are positive.

The second model, proposed in \cite{Miranda:2009rs} without a cosmological constant, is defined by
\begin{eqnarray}
f(R) = R-\alpha R_{*}\ln \left(1+\frac{R}{R_{*}} \right),
\end{eqnarray}
where $\alpha$ and $R_{*}$ are positive parameters. The condition for no ghosts gives: $\alpha< \tilde{R}/R_{*}+1$, where $\tilde{R}$ is the value of the Ricci scalar at the final accelerated fixed point. Since $f(0)=0$, this model also admits a Schwarzschild solution. For a Schwarzschild-de Sitter black hole, the constant curvature $R_0$ from (\ref{tcond}) is given by
\begin{eqnarray}
\label{fc2}
-\frac{\alpha R_{*}R_0}{R_{*}+R_0}=R_0-2\alpha R_{*}\ln \left(1+\frac{R_0}{R_{*}}\right).
\end{eqnarray}
The entropy for the Schwarzschild-de Sitter black hole in this model is obtained from (\ref{31}) as
\begin{eqnarray}
\label{fs2}
S=\left[\frac{(1-\alpha)R_{*}+R_0}{R_{*}+R_0}\right]\pi r^2_+=\left[\frac{(1-\alpha)R_{*}+R_0}{R_{*}+R_0}\right]\frac{A}{4},
\end{eqnarray}
which is $\frac{(1-\alpha)R_{*}+R_0}{R_{*}+R_0}$ times and less than the Bekenstein-Hawking Entropy. For $R_0=0$, the equation (\ref{fs2}) reduces to the entropy of Schwarzschild black hole in this model $S=(1-\alpha)A/4$, also less than the Bekenstein-Hawking Entropy. By computing the energy from (\ref{energy1}), we have
\begin{eqnarray}
\label{fe2}
E&=&\frac{1}{2}\left[\frac{(1-\alpha)R_{*}+R_0}{R_{*}+R_0}\right]r_{+}+\frac{1}{12}\left[\frac{\alpha R_{*}R_0}{R_{*}+R_0}-\alpha R_{*}\ln\left(1+\frac{R}{R_{*}} \right)\right]r^3_{+}\\\nonumber
&=&\frac{1}{2}\left[\frac{(1-\alpha)R_{*}+R_0}{R_{*}+R_0}\right]r_{+}-\frac{1}{24}\left[\frac{(1-\alpha) R_{*}+R_0}{R_{*}+R_0}\right]R_0 r^3_{+}\\\nonumber
&=& \left[\frac{(1-\alpha)R_{*}+R_0}{R_{*}+R_0}\right]M,
\end{eqnarray}
where the conditions (\ref{hc}) and (\ref{fc2}) has been used. For $R_0=0$, the equation (\ref{fe2}) reduces to the energy of Schwarzschild black hole in this model $E=(1-\alpha)M$.
To guarantee the nonnegativity of the entropy and the energy, we must have $\alpha\leq 1+R_0/R_{*}$ for Schwarzschild-de Sitter black hole and $\alpha\leq 1$ for Schwarzschild black hole, giving new constrains on the parameter $\alpha$.

The entropy and the energy of black hole in the above two $f(R)$ theories just are the corresponding values in Einstein's gravity timing the factor $F(r_+)$, compatible with the results obtained in \cite{Dyer:2008hb}. The situation, however, will change for black hole with non-constant Ricci curvature, which we come to discuss in the next subsection.

\subsection{The non-constant Ricci curvature case}
We apply the same procedure for solutions with non-constant curvature which is more interesting but less concerned in literatures. The starting point is the following $f(R)$ theory with the form
\begin{eqnarray}
f(R)=R+2\alpha\sqrt{R},
\end{eqnarray}
where $\alpha$ is a constant ($-\infty<\alpha<0$). This model was found in \cite{Sebastiani:2010kv}, which admits the following solution
\begin{eqnarray}
B(r)=\frac{1}{2}+\frac{1}{3\alpha r},
\end{eqnarray}
with the Ricci scalar $R=1/r^2$. At the horizon, $B(r_+)=0$ gives $r_+=-2/(3\alpha)$. From the equation (\ref{31}), the entropy of the black hole is computed as
\begin{eqnarray}
\label{fs3}
S=\pi r^2_+\left(1+\alpha r_+\right)=\frac{A}{12},
\end{eqnarray}
which is a third of the Bekenstein-Hawking Entropy. The energy of black hole which is derived from (\ref{energy1})
\begin{eqnarray}
E=\frac{1}{8}(4+3\alpha r_+)r_+=-\frac{1}{6\alpha}.
\end{eqnarray}
Since $-\infty<\alpha<0$, guaranteeing that the entropy and the energy both are positive.

The second $f(R)$ model with non-constant curvature we consider here is given by \cite{Amirabi:2015aya}
\begin{eqnarray}
f(R)=R+2\alpha\sqrt{R-4\Lambda}-2\Lambda,
\end{eqnarray}
where $\alpha<0$ and $\Lambda$ is an integration constant which can be interpreted as the cosmological constant. This model allows the solution with $B(r)$ having the form
\begin{eqnarray}
B(r)=\frac{1}{2}+\frac{1}{3\alpha r}-\frac{\Lambda}{3} r^2.
\end{eqnarray}
The Ricci scalar evolves as $R=\frac{1}{r^2}+4\Lambda$. At the horizon, $B(r_+)=0$, which gives
\begin{eqnarray}
\label{f4c}
-\frac{2}{3\alpha}=r_{+}-\frac{2\Lambda}{3} r^3_{+}.
\end{eqnarray}
From the equation (\ref{31}), we get the entropy of the black hole
\begin{eqnarray}
\label{fs4}
S=\pi r_{+}^2\left(1+\alpha r_+\right)=\left(1+\alpha r_+\right)\frac{A}{4},
\end{eqnarray}
which is $1+\alpha r_+$ times and less than the Bekenstein-Hawking Entropy. The nonnegativity of the entropy gives constraint on the parameter $\alpha$: $\alpha\geq-1/r_{+}$. Comparing (\ref{fs3}) and (\ref{fs4}), though the two $f(R)$ models are different, the entropies of black hole in them have the same form, which is interesting. The two values, however, are not the same. From (\ref{31}), the energy is given by
\begin{eqnarray}
E=\frac{1}{24}\left(12+9\alpha r_{+}-4\Lambda r^2_{+}-6\alpha\Lambda r^3_{+}\right)r_+=\frac{1}{12}(3-2\Lambda r^2_+)r_+,
\end{eqnarray}
where the condition (\ref{f4c}) has been used. The nonnegativity of the energy gives constraint on the parameter $\Lambda$: $\Lambda\leq 3/(2r^2_+)$. In these two $f(R)$ theories, the $A$s are not the areas of black holes in Einstein's gravity any more, therefore the entropies also are not the entropies of black holes in Einstein's gravity timing the factor $F(r_+)$. The situation for energy is even more complicated.
\section{Conclusions and discussions}
In this paper the issue whether the new horizon first law still work in $f(R)$ theory has been tackled. We have proposed to identify the entropy and the energy of black hole as quantities proportional to the corresponding value of integration. The identifications are substantiated by the fact that the new horizon first law holds true as a consequence of equations of motion in $f(R)$ theories. The formula for entropy is in agreement with the results obtained by using the Euclidean semiclassical approach or the Wald entropy formula \cite{Dyer:2008hb, Vollick:2007fh, Iyer:1995kg}, and the express of energy is also consistent with that obtained in \cite{Cognola:2011nj} where the entropy was pre-given as that obtained by using the Wald method. For applications, some nontrivial exact black hole solutions in $f(R)$ theories have been considered, the entropies and the energies of black holes in these models are firstly computed, which may be useful for future analysis. In addition, we found that the nonnegativity of the entropy or of the energy can gives new constraints on the parameters of some $f(R)$ theories. The results obtained here, together with other results in literatures, seem to indicate that the thermodynamic origin of a generalized modified gravity, when horizons are present, has a broad validity. Whether the approach presented here is applicable to higher dimensions, rotating black hole, and other higher order gravities remains an interesting subject for future study.

\begin{acknowledgments}
This study is supported in part by National Natural Science Foundation of China (Grant No. 11273010), the Hebei Provincial Outstanding Youth Fund (Grant No. A2014201068), the Outstanding Youth Fund of Hebei University (No. 2012JQ02), and the Midwest universities comprehensive strength promotion project.
\end{acknowledgments}

\bibliographystyle{ieeetr}%{elsarticle-num}
\bibliography{ref}

\end{document}